\documentstyle[prl,aps,preprint,epsfig]{revtex}       
\begin{document}

\draft

\title{Low temperature transport in AC-Driven Quantum Dots in the Kondo regime} 
\author{Rosa L\'opez$^{1}$, Ram\'on Aguado$^{2}$, Gloria Platero$^{1}$ and
Carlos Tejedor$^{3}$}
\address{$^1$ Teor\'{\i}a de la Materia Condensada,
Instituto de Ciencia de Materiales (CSIC) Cantoblanco,28049
Madrid, Spain.}
\address{$^2$Center for Materials Theory, Department of Physics and Astronomy, Rutgers University,
Piscataway New Jersey 08854-8019, USA}
\address{$^3$ Departamento de F\'{\i}sica Te\'orica de la Materia Condensada,
Universidad Aut\'onoma de Madrid, Cantoblanco,28049
Madrid, Spain.}
\date{\today}
\maketitle
\begin{abstract}
We present a fully nonequilibrium calculation of the low temperature 
transport properties of a quantum dot in 
the Kondo regime when
an AC potential is applied to the gate voltage.
We solve a time dependent Anderson model with finite on-site Coulomb interaction.
The interaction self-energy is calculated
up to second order in perturbation theory in the on-site interaction, in the context of the Keldysh 
non-equilibrium technique, and the effect of the AC
voltage is taken into account exactly for all
ranges of AC frequencies and AC intensities.
The obtained linear conductance and time-averaged density of states of the quantum dot evolve  
in a non trivial way as a function of the AC frequency and AC intensity of the harmonic modulation.. 
\end{abstract}

\pacs{PACS numbers: 73.40.Gk, 72.15Qm, 85.30Vw, 73.50.Mx}
\section{Introduction}
Recent experiments \cite{Gold,Leo,Sttut} showing Kondo behavior in the low 
temperature transport of quantum dots (QD's) have opened a new 
arena for the study of strongly correlated electrons in artificial systems. 
Kondo effect in dilute magnetic alloys appears as a crossover from weak 
to strong coupling between delocalized electrons of the host non-magnetic 
metal and the unpaired localized electron of the magnetic impurity as the 
temperature ($T$) is reduced well below the Kondo temperature ($T_K$) 
\cite{Hew}.
This crossover leads to the formation of a singlet state between the 
unpaired localized electron in the impurity and electrons in the host 
metal.\\ 
It is important, however, 
to emphasize the main differences of the Kondo physics in QD's with respect to bulk magnetic impurities.  
The parameters which define the $T_K$ in QD's can be changed in a controlled way
by applying the appropriate combination of gate voltages. So, it is 
possible to study either Kondo or Mixed Valence regimes in the same sample. 
For this to be possible, there is an important requirement: the charging energy 
and level separation of the QD must be significantly larger than the 
level broadening due to the coupling to the leads. 
More importantly, the study of Kondo physics in QD's opens a new road to the study of {\it nonequilibrium} many-body phenomena, 
a relatively young and rich area in contemporary condensed matter physics.\\
In this paper, we address the issue of a QD driven out of equilibrium by means of an AC 
voltage. More specifically, we study theoretically the low temperature
transport properties of a QD with an AC voltage applied to the central gate.
We use a time dependent version of the Anderson model. 
In its simplest formulation, the Anderson 
model, valid for both Kondo
and Mixed Valence regimes in bulk systems, describes a single discrete level with on-site electron-electron 
interaction coupled to a band. The model describes different physical 
regimes which, for QD's, are determined by the following parameters: 
(i) the energy difference between a discrete level in the QD ($\epsilon_0$) 
and the Fermi energy of the leads ($E_F$).
(ii) the tunneling coupling ($\Gamma$) between the discrete level in the 
QD and the electronic states in the reservoirs. (iii) the QD charging energy 
(on-site interaction U), i.e., the energy necessary to add an 
electron to the QD. The relevant energy scale is 
$T_K\simeq\sqrt {2U\Gamma} e^{-\pi[|E_F -\epsilon _0|(U+\epsilon_0)
]/2\Gamma U}$, which is related to the binding energy of the many body state 
\cite{Hew}. For $T<T_K$, the Kondo regime is reached when 
$\epsilon_0<E_F-\Gamma$ and $\epsilon_0>E_F-U+\Gamma$ and 
the Mixed Valence regime is established for $E_F-\Gamma<\epsilon_0
<E_F$ and $E_F-U<\epsilon_0<E_F-U+\Gamma$. 
In the Kondo regime at $T=0$, the low energy excitations (quasi-particles) produce 
a peak at $E_F$ (Kondo resonance or Abrikosov-Suhl resonance) in the density of states (DOS) \cite{Hew}.
One electron at $E_F$ becomes scattered by the QD undergoing a phase 
shift which is proportional to the displaced charge $\delta n$ (Friedel-Langreth 
sum rule (FLSR) \cite{Lan}) and the linear conductance for a QD symmetrically 
coupled to the leads takes the value ${\cal G}=(2e^{2}/h) sin^2(\pi \delta n)$. 
In the symmetric
case ($\epsilon_0=\frac{-U}{2}$), $\delta n=0.5$ leads to a perfect 
transparency of the QD. For any chemical potential between $\epsilon_0$ 
and $\epsilon_0+U$ the QD has the linear conductance as a function of the 
chemical potential of an almost perfectly open
channel $2e^{2}/h$ \cite{Ng,Gla,Kaw}. However, as temperature increases, 
inelastic scattering processes reduce the DOS at $E_F$ (i.e., the 
linear conductance) at the Kondo valley and eventually two peaks 
at $\epsilon_0$ and $\epsilon_0+U$ appear for $T>>T_K$.\\ 
As we 
already mentioned, new questions arise when driving the QD out of 
equilibrium \cite{Hers,Levy,Mei,Win,Ger,time,Integ,Ring,DD,Leo2,Bli,Bru,Hett}. When this is done by means of 
the application of a finite DC voltage bias, the linear conductance is reduced 
and the Kondo peak in the DOS splits\cite{Mei,Win,Ger,time}. 
More sophisticated configurations of QD's in the Kondo regime constitute
a growing area of intense investigations, both from the theoretical and experimental sides.
Time dependent Kondo physics \cite {time}, Kondo physics in integer-spin QD's \cite{Integ}, 
QD's embedded in Aharonov-Bohm rings \cite{Ring} or double QD systems \cite{DD}, 
are examples of such configurations.

We focus on the study of the transport properties of an Anderson 
Hamiltonian with a time dependent resonant level $\tilde{\epsilon}_0(t)=
\epsilon_0+V_{AC}cos(\omega_0 t)$. This can be achieved experimentally by 
means of a time dependent central gate voltage capacitively coupled to the 
QD. In the high temperature regime these type of experiments have indeed 
been carried out leading to the observation of photon-assisted processes in 
the Coulomb blockade regime \cite{Leo2,Bli}. This regime has been studied 
theoretically as well \cite{Bru}. In the same way, there has been some 
theoretical effort devoted to studying the AC transport at very low ($T<<T_K$) 
and intermediate temperatures ($T\lesssim T_K$) in the Kondo and 
Mixed Valence regimes \cite {Hett,Ng2,Sch,Win2,Ros,Avis,Gla2}.
Recently transport in an AC driven quantum dot at low temperatures has
been measured as well\cite{Sil}.

In this work we clarify the role of an AC voltage in the Kondo effect in QD's. 
We concentrate on the low temperature regime so that a Fermi liquid 
theory is adequate in spite of the fact that it
overestimates the width of the Kondo peak, i.e., the $T_K$ 
\cite{Hew,Noz}. In fact our Fermi liquid approach 
gives a resonance width which decreases algebraically with U, instead of 
having an exponential decay as given by scaling calculations
\cite{Hew}. The occupation of the QD as well as all the relevant 
quantities in transport have to be 
calculated by using non-equilibrium propagators. Finally, some approximation 
is needed for calculating the Green's function of the QD. We use a finite-U 
perturbation theory for the Green's function of the impurity. The Anderson 
Hamiltonian is solvable through the Bethe Ansatz \cite{Beth}, Numerical 
Renormalization Group \cite{Cos} and Quantum Monte Carlo methods \cite{MC} 
but a reliable and simple method to obtain dynamical properties at low 
temperatures in the whole range of
interactions ($U/\Gamma$) is not available. Previous efforts have concentrated 
on the $U\rightarrow\infty$ limit where a Non-Crossing
Approximation (NCA) \cite{Mei,Win,Hett,Win2} can be made for high and intermediate 
temperatures. However, such approximations do not give
the exact local Fermi-liquid properties as $T\rightarrow 0$. They
cannot describe the transition from the weak-correlation to the 
strong-correlation regime. On the other hand, finite-$U$ perturbation 
theory \cite{Hers,Yos,Hor} describes the symmetric case properly, but 
presents clear anomalies away from this special situation (which can be overcome by means of an interpolative self-energy \cite {Levy,Ros}). 
In this paper we will restrict ourselves to the symmetric case which is
the relevant one for the experimental information available \cite{Sil}.
The study of the asymmetric case will be analyzed elsewhere.

The main difficulty for our purpose resides in the determination of 
the QD Green's function, and specifically of its self-energy. In 
a previous paper\cite{Ros} we proposed an ansatz for the modification of the
QD Green's function due to an harmonic modulation. 
Here, we improve our previous description, valid in the limit of 
small interaction U, and extend the calculation to finite temperatures.

The paper is organized as follows:
in Sec. II we describe the theoretical model and deduce
the expressions for the self-energy and the time-averaged spectral density. 
In Sec. III we present the results for the time-averaged density of states and
the linear conductance at finite temperatures and for different AC frequencies 
and AC intensities of the harmonic modulation. Moreover, we compare with previous experiments and theoretical models. Finally, Sec. IV summarizes the main 
conclusions of the paper.  

\section{Theoretical Model}
\subsection{Keldysh Formalism}
\begin{figure}
\centerline{
\psfig{figure=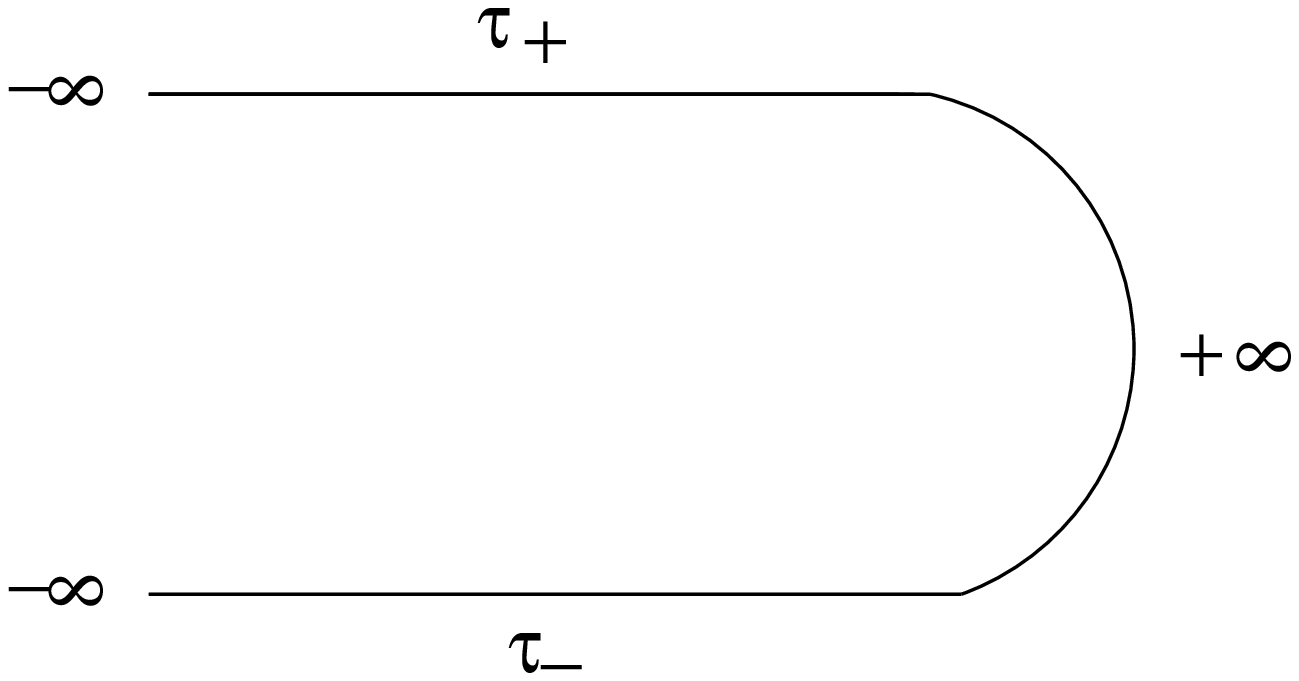,width=8.0 cm, height=5.0 cm}}
\caption{Complex time-contour. 
The times in the positive branch are $\tau_+$ while times in the negative branch are $\tau_-$.}
\label{fig:contour}
\end{figure}

The application of a time dependent component to the energy level in the Anderson Hamiltonian (See Eq. 6, below) breaks the time translational invariance of the system and, then, we need an
approach capable of addressing this fully non-equilibrium situation. When the time dependent perturbation acts for a while, the system does not recover its thermodynamic equilibrium after the
perturbation is over. The whole process has not the symmetry between $t\rightarrow 
-\infty$ and  $t\rightarrow \infty$ and, then, it is not 
possible an equilibrium expansion in terms of expectation values. 
Nevertheless, the problem can be solved by allowing the system to evolve 
from $-\infty$ to the moment of interest $\bar t$ and then continuously 
evolve from $t=\bar t$ back to $t\rightarrow -\infty$. In this way 
all the expectation values are evaluated in a well-defined state 
which was prepared in a remote past. This special complex time-contour 
is the main ingredient of the non-equilibrium Keldysh formalism 
\cite{Keldysh}. By using this complex time-contour (see Fig. 
\ref{fig:contour}) the formal expression for the Green's function 
relative to a total Hamiltonian $H=H_0+V_{int}$ which includes 
an interaction potential $V_{int}$ is given by
\begin{eqnarray}\label{GF}
G_{d,\sigma}(t,t^\prime)=\langle T_c [S_-(-\infty, \infty)S_+(\infty, -\infty)d_\sigma(t), d_\sigma^\dagger(t^\prime)]\rangle,
\end{eqnarray}
where $S_+(\infty, -\infty)$ is the usual S-matrix, 
$exp[-i\int_{-\infty}^{\infty}V_{int}(\tau)d\tau]$ and 
$S^\dagger(\infty, -\infty)=S(-\infty, \infty)$. 
$d_\sigma(t)$,  $d_\sigma^\dagger(t^\prime)$ are operators 
in the interaction picture. $S_{+}(\infty, -\infty)$ depicts the causal branch 
(from $-\infty$ to $\infty$), while $S_{-}(-\infty, \infty)$ depicts 
the anticausal branch (from $\infty$ to $-\infty$). The Green's function defined 
in Eq. (\ref{GF}) is a matrix $2\times 2$, since both times, $t$ and $t'$ 
can belong to the causal branch (causal Green's function, $G_{d,\sigma}
^C(t,t')$ ) or to the anticausal one (anticausal Green's function, $\bar{G}
_{d,\sigma}^C(t,t')$) and it is possible to have one time in the causal 
branch and the other one in the anticausal branch (which defines the lesser, 
$G_{d,\sigma}^<(t,t')$ and greater $G_{d,\sigma}^>(t,t')$ Green's functions). 
$T_c$ is the contour-ordering operator which arranges the operators on the 
closed time-path in this way: operators with time
labels later on the contour are moved left of operators
of earlier time labels. In this way, the non-equilibrium perturbation theory 
has the same structure as the equilibrium perturbation theory keeping in mind 
the label of the times. The $2\times 2$ matrix for the Green's function reads 
\begin{eqnarray}        
G_{d\sigma}(t,t') \equiv \left( \begin{array}{cc}          
G_{d,\sigma}^C(t,t')  &  G_{d,\sigma}^<(t,t')   \\                     
G_{d,\sigma}^>(t,t') & \bar G_{d,\sigma}^C(t,t')             
\end{array} \right),  
\end{eqnarray}
where 
\begin{eqnarray}\label{gnkalpha}
&&G_{d,\sigma}^C(t,t')  \equiv -i\langle {\cal{T}} \{{d}_{\sigma}(t),
{d}^{\dagger}_\sigma(t')\}\rangle,\;\;
\bar{G}_{d,\sigma}^C(t,t')  \equiv -i\langle \bar{\cal{T}} \{{d}_{\sigma}(t)
,{d}^{\dagger}_\sigma(t')\}\rangle,\nonumber \\
&&G_{d,\sigma}^<(t,t')  \equiv i\langle d^{\dagger}_{\sigma}(t') d_\sigma(t)\rangle,\;\;
G_{d,\sigma}^>(t,t') \equiv -i\langle{d}_\sigma(t) d^{\dagger}_{\sigma}(t')\rangle.
\end{eqnarray}   
Here, ${\cal{T}}$ is the causal time-ordering operator, and $\bar{ \cal{T}}$ 
is the anticausal time-ordering operator.
In the conventional Keldysh matrix formulation of the perturbation theory, 
one does not work directly with the Green's function defined on the 
complex time-contour, but with a linear combination of these four possible 
time orders. The usual linear combinations are (similar relations hold for 
the self-energies)
\begin{eqnarray}\label{ret_and_adv}
&&G_{d,\sigma}^r(t,t')=\theta(t-t')[G_{d,\sigma}^>(t,t')-G_{d,\sigma}^<(t,t')],
\nonumber\\
&&G_{d,\sigma}^a(t,t')=\theta(t'-t)[G_{d,\sigma}^<(t,t')-G_{d,\sigma}^>(t,t')],
\end{eqnarray}
where $G_{d,\sigma}^r(t,t')$ is the retarded Green's function, $G_{d,\sigma}^a(t,t')$ is the advanced Green's function,
and $G_{d,\sigma}^<(t,t')$, $G_{d,\sigma}^>(t,t')$ are the so-called lesser and greater Green's functions, respectively.
\subsection{Hamiltonian}
The time dependent Anderson Hamiltonian is:
\begin{eqnarray}\label{Hamiltonian}
&&H  =  H_{leads} + H_{QD} + H_{sd}+ H_{AC}(t),
\end{eqnarray}
where
\begin{eqnarray}\label{Hamiltonian1}
&&H_{leads}  = \sum _{ k\in \{ L,R \}, \sigma} \epsilon _k c^\dagger _{ k, 
\sigma},
c_{ k,\sigma},
\nonumber \\
&&H_{QD} =  \sum _\sigma \epsilon _{0,\sigma} d^\dagger_\sigma d_\sigma + U d^\dagger_\uparrow d_\uparrow
d^\dagger_\downarrow d_\downarrow, 
\nonumber \\
&&H_{sd} =  \sum _{k\in \{L,R\},\sigma }V_k c^\dagger _{k, \sigma} d_\sigma
+V^*_k d^\dagger_\sigma c_{k, \sigma}, 
\nonumber \\
&&H_{AC}(t) =  \sum _\sigma V_{AC}cos \omega_0 t \; d^\dagger_\sigma d_\sigma. 
\end{eqnarray}
$ V_{AC}$ and $\omega_{0}$ are the AC intensity and AC frequency of the
AC potential respectively.
$d^\dagger_\sigma$ creates an electron with spin $\sigma$ in the QD, while 
$c^\dagger _{k, \sigma}$ creates it in the lead with energy $\epsilon_k $ 
($k$ labels the rest of quantum numbers). The AC voltage
modulates in time the relative position of the QD level $\epsilon _{0,\sigma} $
with respect to $E_F$. An eventual breakdown of the spin
degeneracy would be represented by $\epsilon _{0,\sigma}\neq \epsilon 
_{0,\bar{\sigma}}$.
The coupling $V_{k}$ between the QD and the leads produces a broadening
$\Gamma^{L(R)}(\epsilon)=-2Im [ \Sigma^{L(R)}_{sd} (\epsilon +i \delta )]=2
\pi\sum_{k\in L(R)}|V_{k} |^2\delta(\epsilon-\epsilon_{k})$, where 
$\Sigma^{L(R)}_{sd}$ is the hybridization single-particle self-energy. 
Hereafter, for simplicity, we consider the wide band (WB) limit approximation 
which neglects the principal value of the hybridization self-energy 
and considers the imaginary part to be an energy independent constant, i.e., 
$\Sigma^{L(R)}_{sd}(\epsilon)=\Lambda^{L(R)}(\epsilon)-i\Gamma^{L(R)}(\epsilon)/2\approx-i\Gamma^{L(R)}/2$. 
\subsection{Model}
Here, we discuss the procedure for obtaining the QD Green's 
functions which 
allows us to obtain the spectral density of the QD and the linear conductance.\\ 
In the remote past, the QD is decoupled from the leads. The coupling between 
different regions (the contacts and the central region) is treated as a 
perturbation by means of standard equilibrium perturbation theory. In a 
first step, the effect of the on-site interaction is included via a Hartree 
mean-field approximation (see Appendix A). 
The time modulation 
of the QD level is treated via non-equilibrium perturbation theory, 
since the time translational invariance is broken by the AC voltage.
At this point, we include the 
correlation effects by computing the on-site interaction self-energy (lesser and greater) up to 
second order by means of the diagrams of Fig. \ref{fig:diagrama}. 
These diagrams are evaluated by using the previous lesser and greater Green's functions 
as bare propagators (Appendix B). These bare propagators include the coupling between the QD and the 
contacts, the time dependence of the QD level and the on-site interaction 
in the Hartree approximation. Once the correlation self-energy has 
been calculated, the QD retarded Green's function is obtained by means of 
the time dependent Dyson equation (Eq. 13). Finally, the time-averaged spectral density (Eq. 21) 
and the linear conductance (Eq. 22)
are calculated from the QD retarded Green's function.
\subsection{Correlation self-energy}
\begin{figure}[h]
\centerline{
\psfig{figure=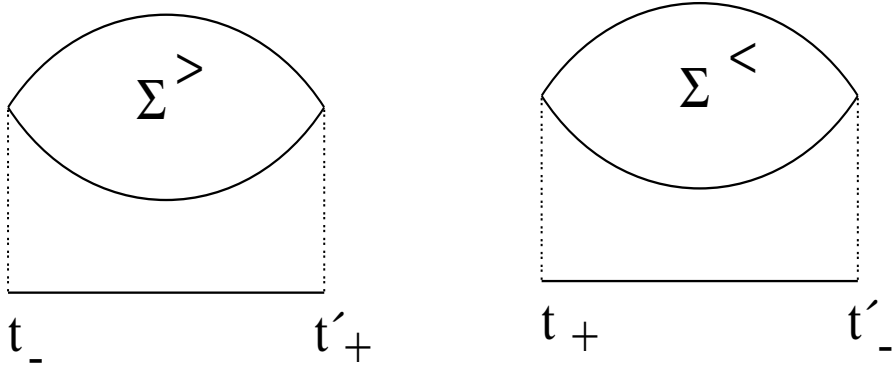, width=12 cm, height=5 cm}}
\caption{Self-energies of order $U^2$, $\Sigma^{>,2}_{d,\sigma}(t,t')$ and $\Sigma^{<,2}_{d,\sigma}(t,t')$. 
The times in the causal branch are marked with a $+$ symbol whereas the times in the anticausal branch are marked
with a $-$ symbol.
Solid lines denote QD Green's functions in the Hartree approximation 
including coupling to the leads and AC potential. Dashed lines correspond to the on-site repulsion U electron-electron interaction.}
\label{fig:diagrama}
\end{figure}

The starting point for the derivation of the correlation self-energy in 
the presence of the AC potential, consists of calculating the 
lesser and greater QD Green's functions, in the Hartree 
approximation, including coupling to the leads.
In the absence of the time dependent potential, the retarded and 
advanced QD Green's functions have the following expressions (see Appendix A)
\begin{eqnarray}
{\mathbf {g}}^{r,a}_{d,\sigma}(t-t')= \mp i\theta(\pm t \mp t')
\exp^{-i\int_{t'}^{t}dt_1(\epsilon_{0,\sigma}+Un_{d,\bar{\sigma}}\mp i\sum_{\alpha\in L,R}\Gamma_\alpha/2)}\;,
\end{eqnarray}
$n_{d,\sigma}=\langle d_\sigma^\dagger(t)d_\sigma(t)\rangle$ being the QD 
occupation. Note that these QD Green's functions have been calculated taking 
into account the coupling self-energy (which is given by $\mp i\sum_
{\alpha\in L,R}\Gamma_\alpha/2$) and the interaction in the Hartree 
approximation (given by $Un_{d,\bar{\sigma}}$).
Now, if one also considers a time modulation of the QD level, the retarded 
and advanced QD Green's functions have the following forms (see Appendix B)
\begin{eqnarray}\label{ret_adv_GF}
G^{r,a}_{d,\sigma}(t,t')=e^{-i\frac{V_{AC}}{\omega_0}
(sin\omega_0 t-sin\omega_0 t')}{\mathbf {g}}^{r,a}_{d,\sigma}(t-t').
\end{eqnarray}
The lesser and greater QD Green's functions can be obtained through the 
well known relation \cite{Jau,Lan2}
\begin{eqnarray}\label{less_grea_GF}
G^{<,>}_{d,\sigma}(t,t')=\int dt_1 \int dt_2 G^r_{d,\sigma}(t,t_1)\Sigma^{<,>}_{sd}(t_1,t_2)G^a_{d,\sigma}(t_2,t'),
\end{eqnarray}
where $\Sigma^{<,>}_{sd}(t_1,t_2)$ are the lesser and greater 
coupling self-energies defined in Appendix A (Eqs. A2 and A3, respectively).
Now we include correlation effects up to second order in the on-site Coulomb interaction (see Fig. \ref{fig:diagrama}).
The new lesser and greater correlation self-energies  
are calculated by means of the diagrams of Fig. \ref{fig:diagrama} with bare lines which are given by the 
propagators of Eq. (\ref{less_grea_GF}) (analytical expressions are given in Appendix B, Eqs. B11-B14):
\begin{eqnarray}\label{int_self>}
\Sigma^{>,(2)}_{d,\sigma}(t,t')=iU^2G^>_{d,\sigma}(t,t')G^<_{d,\bar{\sigma}}(t',t)G^>_{d,\bar\sigma}(t,t'),
\end{eqnarray}
and 
\begin{eqnarray}\label{int_self<}
\Sigma^{<,(2)}_{d,\sigma}(t,t')=-iU^2 G^<_{d,\sigma}(t,t')G^>_{d,\bar{\sigma}}(t',t)G^<_{d,\bar\sigma}(t,t').
\end{eqnarray}  
The retarded self-energy
\begin{eqnarray}\label{int_self}
\Sigma^{r,(2)}_{d,\sigma}(t,t')=\theta(t-t')[\Sigma^{<,(2)}_{d,\sigma}(t,t')-\Sigma^{>,(2)}_{d,\sigma}(t,t')],
\end{eqnarray}
is obtained from Eqs. (\ref{int_self>}) and (\ref{int_self<}).
\subsection{Dyson equation}
The next step for deriving the retarded QD Green's function consists of 
solving the retarded time dependent Dyson equation. By using Eq. (\ref{int_self})
for the self-energy, one can write
\begin{eqnarray}\label{Dyson_equation}
\biggl[i\frac{\partial}{\partial t}-(\bar{\epsilon}_{0,\sigma}(t)-i\sum_{\alpha\in L,R}\Gamma_\alpha/2)\bigg]{\mathbf G}^{r,(2)}_{d,\sigma}(t,t')=\delta(t-t')+\int dt_1 \Sigma^{r,(2)}_{d,\sigma}(t,t_1){\mathbf 
G}^{r,(2)}_{d,\sigma}(t_1,t'),
\end{eqnarray}
where $\bar{\epsilon}_{0,\sigma}(t)$=$\epsilon_{0,\sigma}+Un_
{d,\bar{\sigma}}(t)+V_{AC}cos \omega_0t$. Eq. (\ref{Dyson_equation})
simplifies considerably by making the gauge transformation 
\begin{eqnarray}\label{gauge_1}
{\mathbf G}^{r,(2)}_{d,\sigma}(t,t')=-i\theta(t-t')e^{-i\int_{t'}^t d\tau 
(\bar{\epsilon}_{0,\sigma}(\tau)-i\sum_{\alpha\in L,R}\Gamma_\alpha/2)}\bar{\mathbf {g}}_{\sigma}(t,t').
\end{eqnarray} 
In the presence of time modulation, the retarded Dyson equation becomes
\begin{eqnarray}\label{reduced_Dyson}
\frac{\partial}{\partial t}\bar{\mathbf {g}}_{\sigma}(t,t')=-\int_{t'}^{t} 
dt_1 K_\sigma (t,t_1)\bar{\mathbf {g}}_{\sigma}(t_1,t'),
\end{eqnarray}
which is defined only for $t\geq t'$ due to the $\theta$ function appearing 
in Eq. (\ref{gauge_1}). $K_\sigma (t,t')$ is the kernel of the 
integro-differential time dependent Dyson equation which is related to the retarded self-energy through the relation:
\begin{eqnarray}\label{Kernel}
\Sigma^{r,(2)}_{d,\sigma}(t,t_1)=-i\theta(t-t_1)K_\sigma(t,t_1)\times 
e^{-i{\int_{t_1}^{t}} d\tau (\bar{\epsilon}_{0,\sigma}(\tau)-i\sum_{\alpha\in L,R}\Gamma_\alpha/2)}.
\end{eqnarray}
When $t=t'$, an additional condition must be imposed in Eq. (\ref{Dyson_equation}):
${\mathbf G}^{r,(2)}_{d,\sigma}(t,t)=-i\langle 
\{d_\sigma(t),d_\sigma^\dagger(t)\}\rangle=-i$, where $\{\}$ is the anticonmmutator. 
This condition implies that the solution of Eq. (\ref{reduced_Dyson}) 
when $t=t'$ is $\bar{\mathbf {g}}_{\sigma}(t,t)=1$.

We solve Eq. (\ref{reduced_Dyson}) by discretizing 
the temporal variables, the partial derivative is replaced by the finite difference
\begin{eqnarray}\label{partial}
\frac{\partial}{\partial t}\bar{\mathbf g}_{\sigma}(t,t')\rightarrow \frac{\bar{\mathbf g}_{\sigma}(m,n)-\bar{\mathbf g }_{\sigma}(m-1,n)} {\delta},
\end{eqnarray}
where $\delta$ is the grid spacing in time space. The integral is converted into a sum 
\begin{eqnarray}\label{integral}
-\int_{t'}^{t} 
dt_1 K_\sigma (t,t_1)\bar{\mathbf {g}}_{\sigma}(t_1,t')\rightarrow -\delta\sum _{k=n}^m
c_k K_\sigma (m,k)\bar{\mathbf {g}}_{\sigma}(k,n) .
\end{eqnarray}
Now, the indexes  $m, n$  and $k$ replace the time arguments 
which appear in the (Eq. \ref{reduced_Dyson}). The coefficients $c_k$ are 
equal to 1 except when $m=k$ or $n=k$ in which $c_k=\frac{1}{2}$.
In this way, the discretized time dependent Dyson equation has the following form,
\begin{eqnarray}\label{discretized}
\bar{\mathbf {g}}_{\sigma}(m,n)=&&\bar{\mathbf {g}}_{\sigma}(m-1,n)-\delta^2 
\sum _{k=n}^m
c_k K_\sigma (m,k)\bar{\mathbf {g}}_{\sigma}(k,n).
\end{eqnarray}
Eq. (\ref{discretized}) constitutes a set of linear equations that can be solved by standard numerical techniques.
Its solution gives the retarded QD Green's function which is used to study the transport properties of the system in 
the next section.
\subsection {Time-averaged spectral density and Linear Conductance}
The time dependent spectral density $\rho_{\sigma}(\epsilon, \bar{t})$, being $\bar{t}=\frac{t+t'}{2}$, is defined as 
the imaginary part of the Fourier transform with respect to $\tau =t-t'$ of the 
retarded QD Green's function
\begin{equation}\label{DOS}
\rho_\sigma(\epsilon,\bar{t})= -\frac{1}{\pi} Im\int_{-\infty}^\infty {\mathbf 
G}^{r,(2)}_{d,\sigma}(\bar{t}+\tau/2,\bar{t}-\tau/2)e^{i\epsilon\tau}d\tau.
\end{equation}
 Since the measurement of the linear conductance implies a time-average 
in $\bar{t}$, we work with the time-averaged spectral density which reads
\begin{equation}\label{time_DOS}
\langle A_{\sigma}(\epsilon)\rangle =\frac{\omega_0}{2\pi}\int_{0}^{\frac{2\pi} {\omega_0}}d\bar{t} \rho_\sigma(\epsilon,\bar{t}).
\end{equation}
The linear conductance \cite{Jau} at finite temperature, 
in terms of the time-averaged spectral density, is given by
\begin{eqnarray}\label{conductance}
{\cal G} &=&\frac{e^2}{\hbar}
\int 
d\epsilon \, \frac{\Gamma_L\Gamma_R}{\Gamma_L+\Gamma_R}
\left( -\frac{\partial f(\epsilon)}{\partial \epsilon} \right)
\sum_\sigma \langle A_{\sigma}(\epsilon)\rangle, 
\end{eqnarray}
where $f(\epsilon)$ is the Fermi-Dirac distribution function.

\section{Results}
\begin{figure}[h]
\centerline{
\psfig{figure=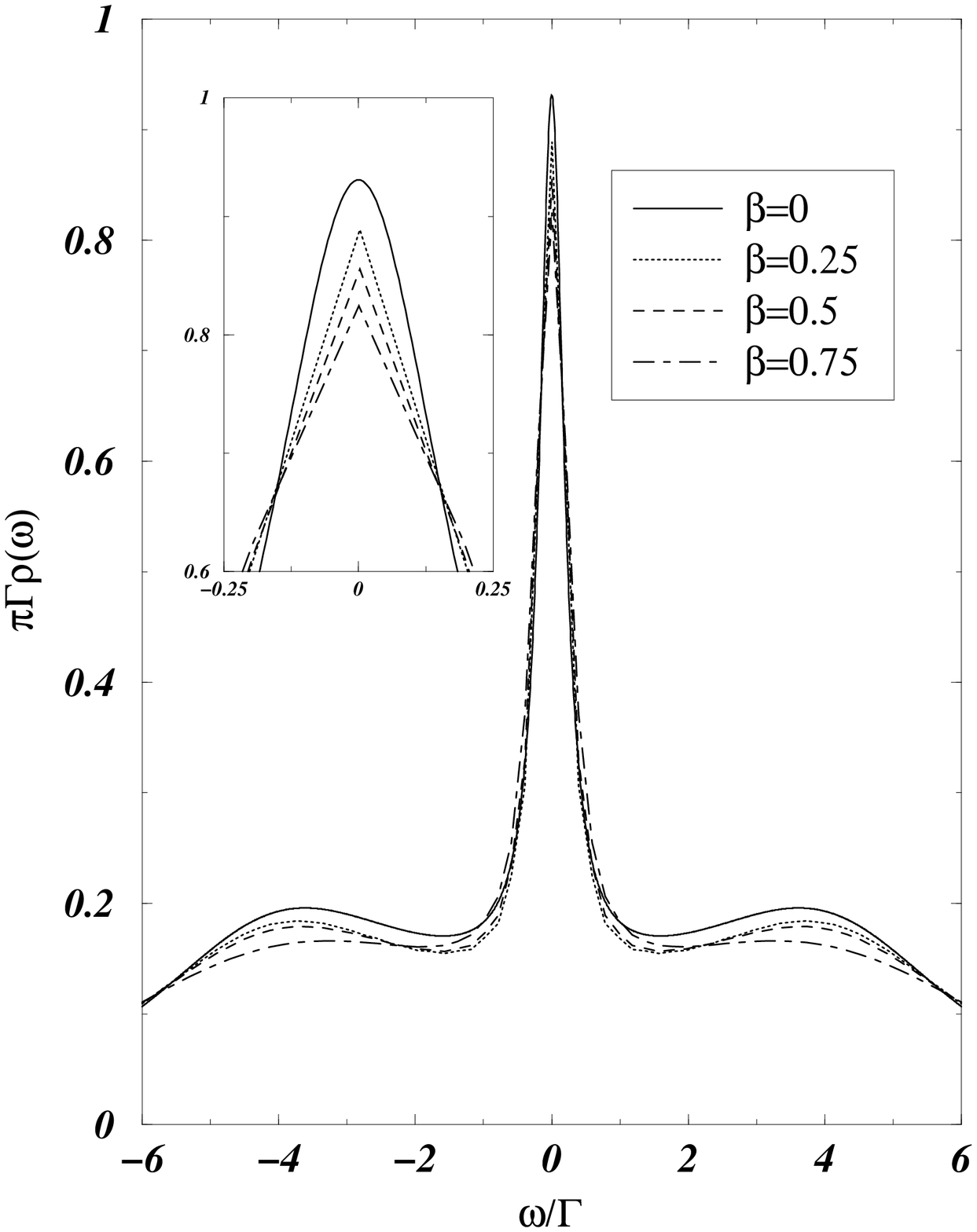, width=0.3\textwidth}
\psfig{figure=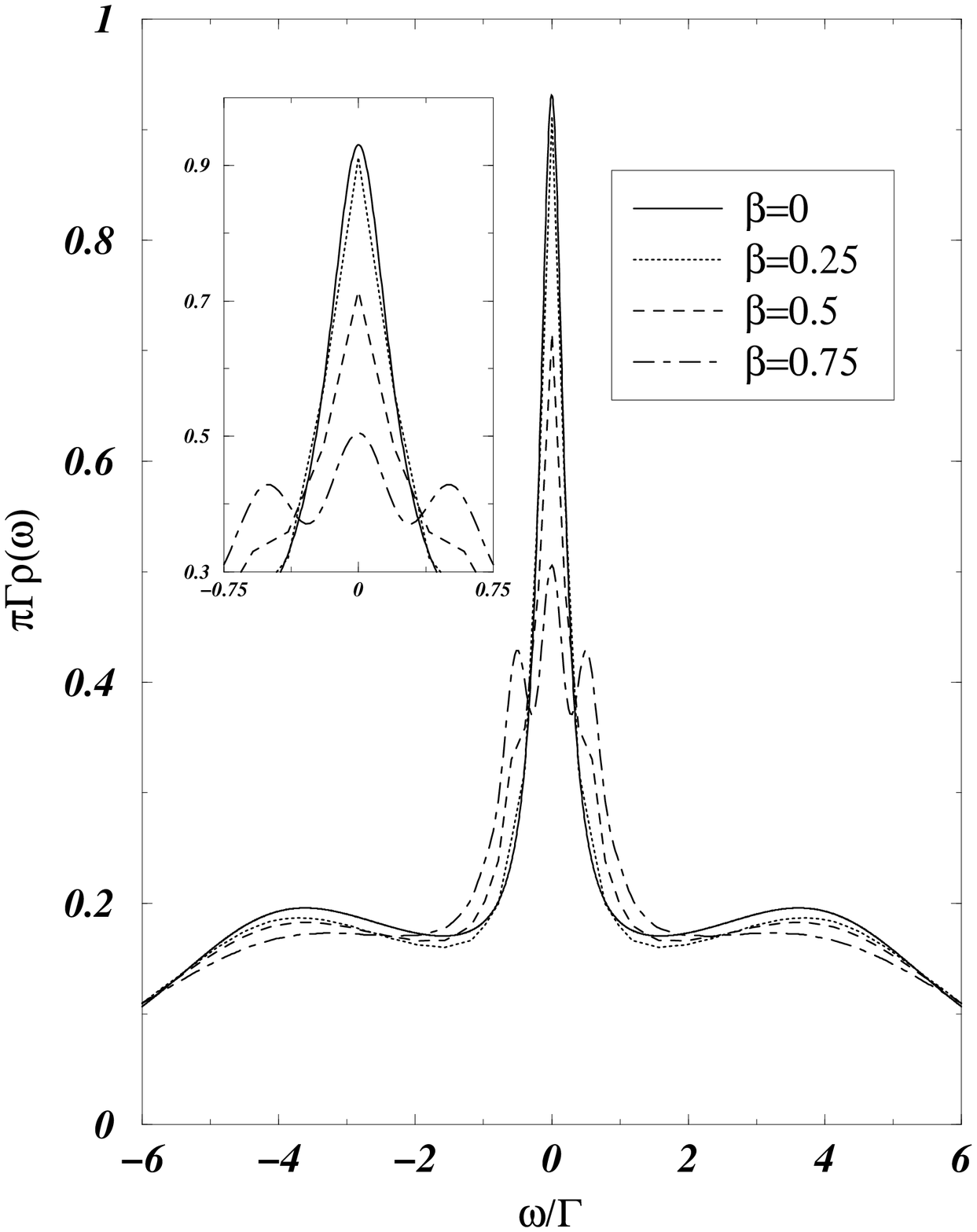, width=0.3\textwidth}
\psfig{figure=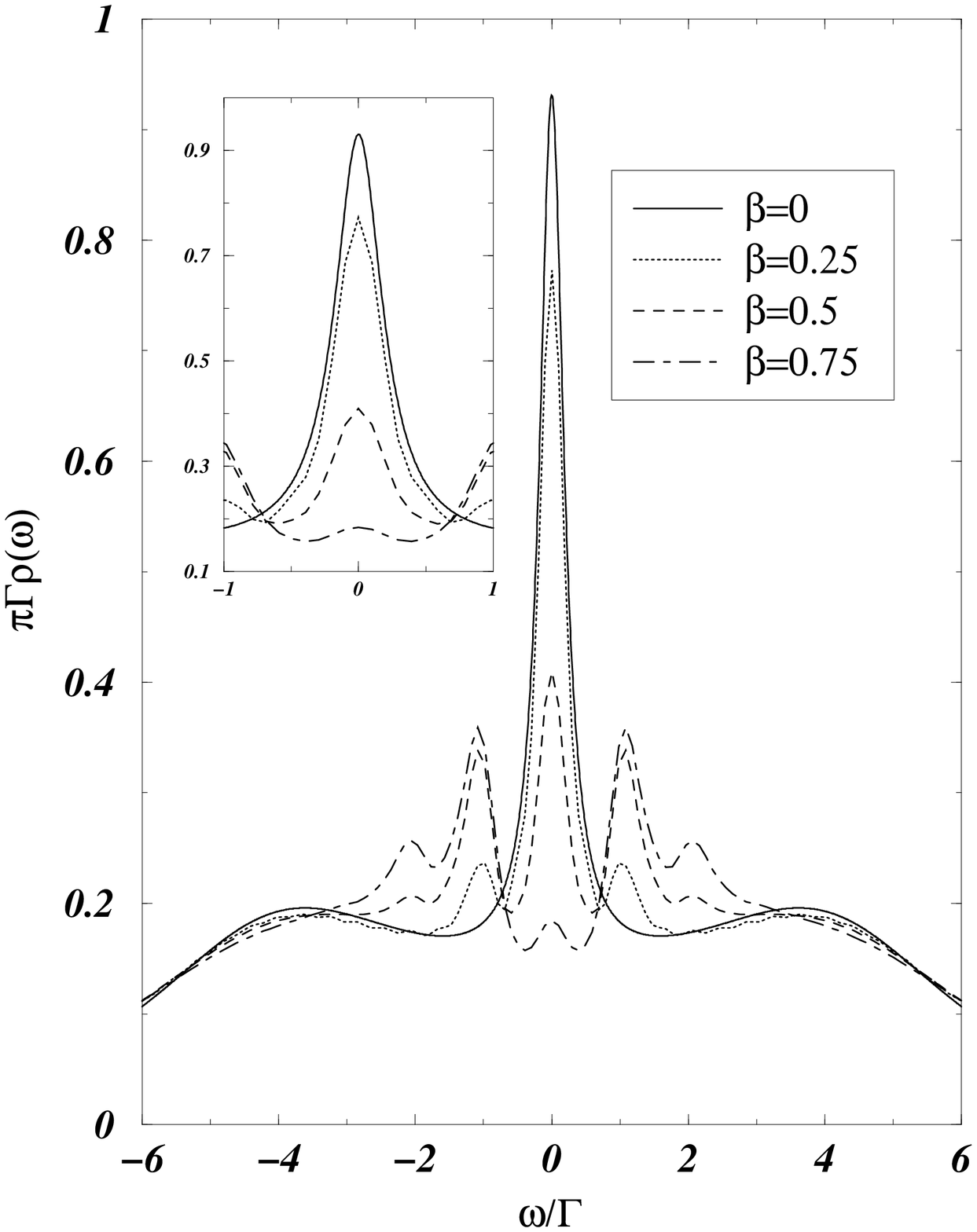, width=0.3\textwidth}
}
\caption{Time-averaged DOS in the strongly correlated regime, 
$U=2.5\pi\Gamma$ and $T=0.05\Gamma$.
(a) The AC frequency is $\omega_0=\Gamma/4\approx 2T_K$, the solid 
line is the case in the absence of the AC potential, i.e., $\beta=0$, where 
$\beta=\frac{V_{AC}}{\omega_0}$. In this case the peak at $E_F$ reaches a 
height of 0.93. The 
dotted line corresponds to $\beta=0.25$ ($V_{AC}=0.0625\Gamma\approx T_K/2$).
The dashed line shows the case of 
$\beta=0.5$ ($V_{AC}=0.125 \Gamma \approx T_K$).
The dot-dashed line corresponds to 
$\beta=0.75$ ($V_{AC}=0.1875\Gamma\approx 3T_K/2$). 
In the three cases the Kondo peak is slightly reduced by the AC signal.\\
(b) $\omega_0=\Gamma/2\approx 4T_K$. The solid line corresponds to $\beta=0$,
the dotted line 
to $\beta=0.25$ ($V_{AC}=0.125\Gamma \approx T_K$). 
The dashed line shows the case of 
$\beta=0.5$ ($V_{AC}=0.25\Gamma\approx 2T_K$) where the peak has
 been significantly reduced and 
the dot-dashed line shows the case of an intense signal where $\beta=0.75$ 
($V_{AC}=0.375\Gamma\approx 3T_K$). In this case the replicas of the Kondo become 
apparent (located at $\omega_0$ and $-\omega_0$) and the peak at 
$E_F$ has been strongly reduced. Figure (c) corresponds to
 $\omega_0=\Gamma$:  the solid line is for the case $\beta=0$; the 
dotted line 
shows the case $\beta=0.25$ ($V_{AC}=0.25\Gamma\approx 2T_K$). At this 
AC intensity the first satellites of the Kondo resonance show up. The dashed line 
corresponds to $\beta=0.5$ ($V_{AC}=0.5\Gamma\approx 4T_K$), in this
case the time-averaged DOS at $E_F$ 
has been suppressed below 0.5. The 
dot-dashed line corresponds to a very 
intense signal, $\beta=0.75$ ($V_{AC}=0.75\Gamma\approx 6T_K$) , 
the Kondo peak in this case
has completely disappeared. Insets: Blow up of the time-averaged DOS near $\omega=0$.}
\label{fig:1}
\end{figure}

\begin{figure}[h]
\centerline{
\psfig{figure=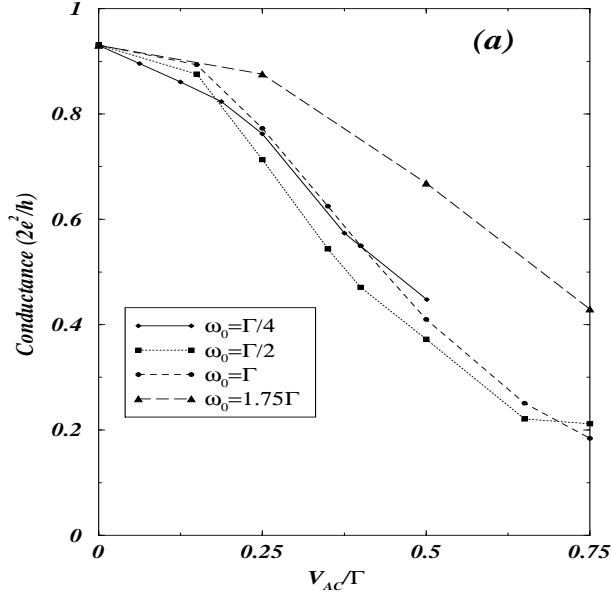, height=8.0cm, width=8.0 cm}}
\caption{ (a) Conductance as a function of $V_{AC}$ in the 
correlated regime where, $U=2.5\pi\Gamma$ and
$T=0.05\Gamma$ for four AC frequencies.}
\label{fig:2}
\end{figure}

\begin{figure}[t]
\centerline{
\psfig{figure=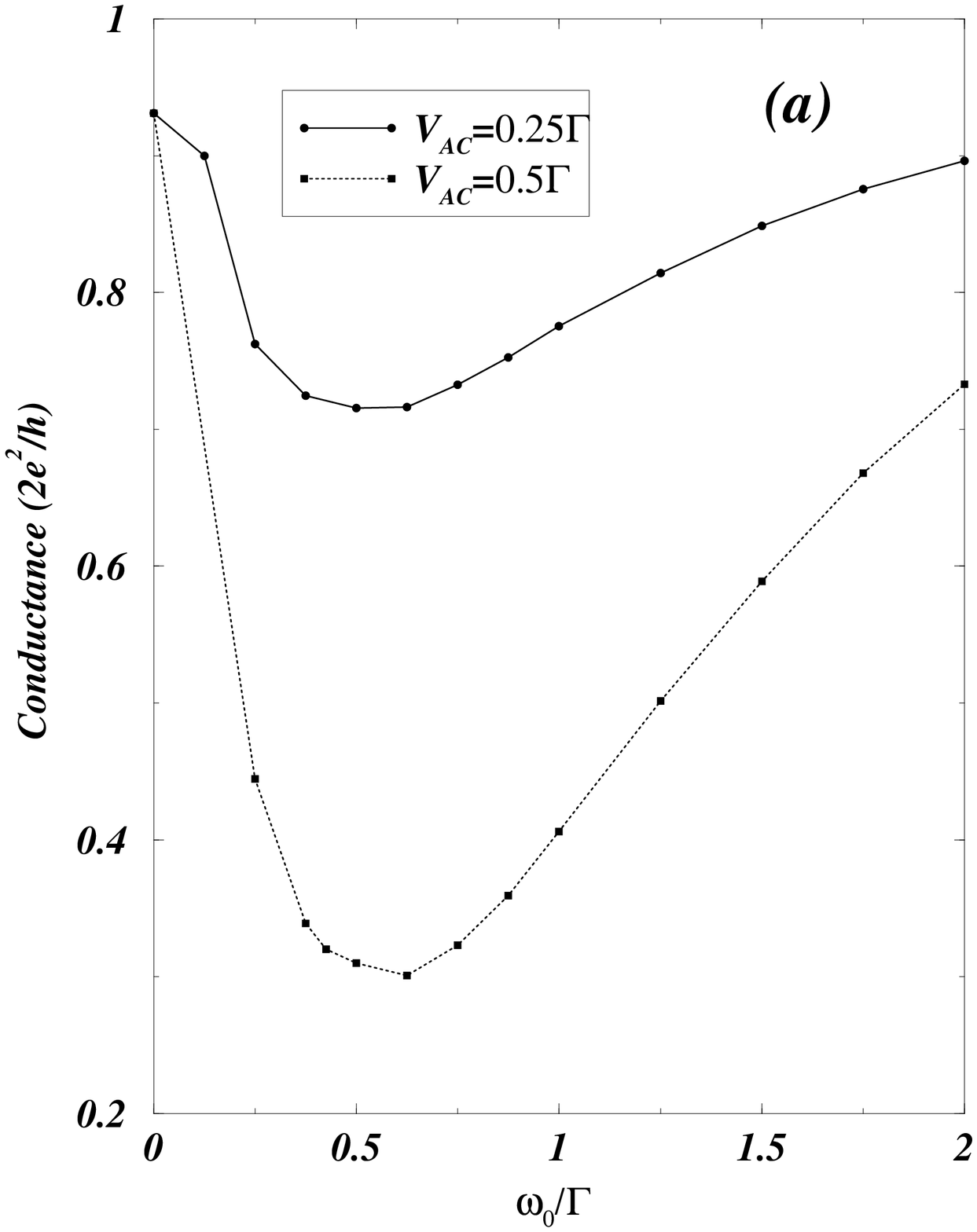, height=8.0cm, width=8.0 cm}
\psfig{figure=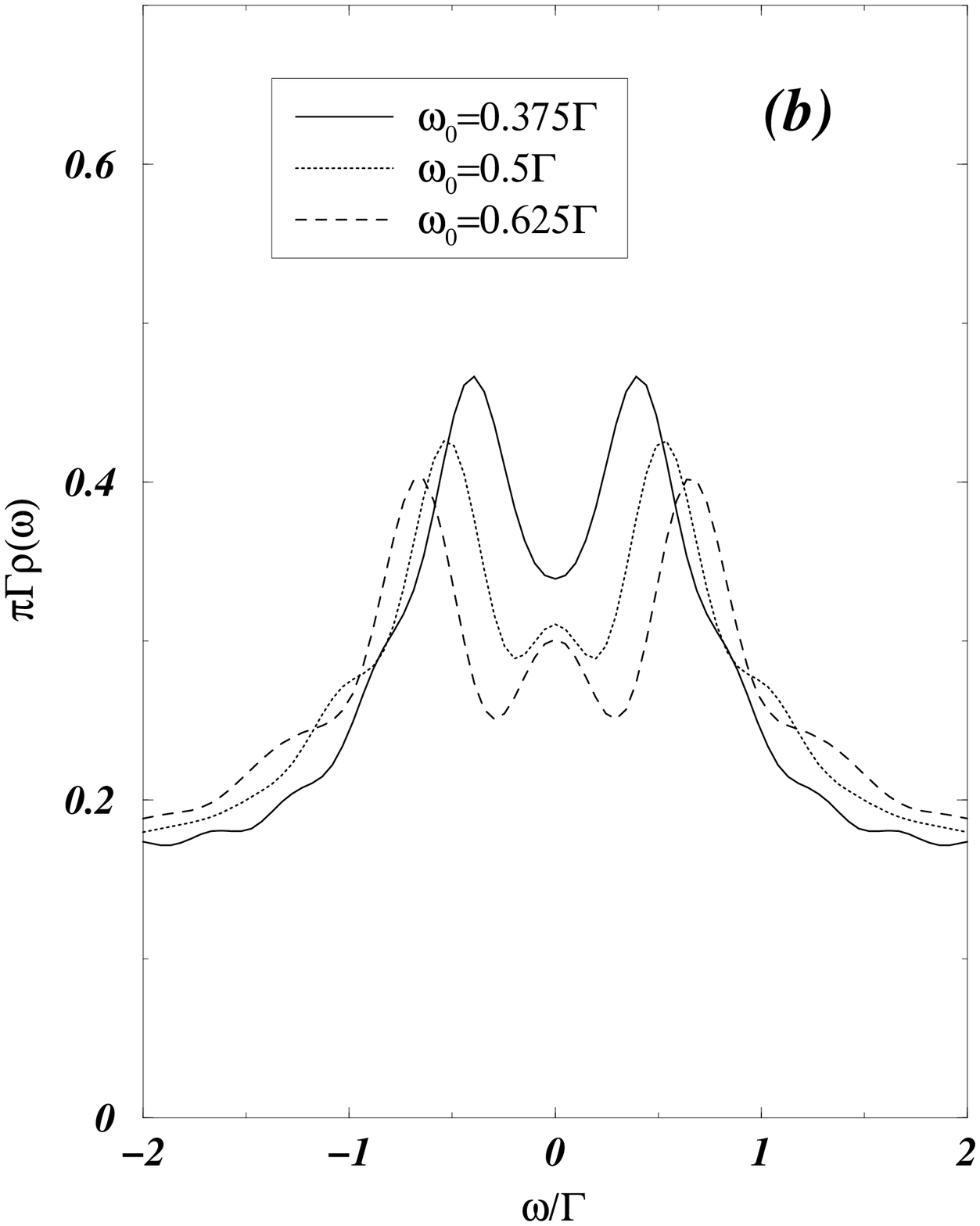, height=8.0cm, width=8.0 cm}}
\caption{(a) Conductance as a function of $\omega_{0}$ 
in the strong correlated regime $U=2.5\pi\Gamma$ and at $T=0.05\Gamma$. The 
solid line shows the case of $V_{AC}=0.25\Gamma\approx 2T_K$,
the dotted line corresponds to the case of stronger AC intensity 
$V_{AC}=0.5\Gamma\approx 4T_K$ (b) Time-averaged DOS for the cases
 of $\omega_0=0.375\Gamma$ (solid line), $\omega_0=\Gamma/2$ (dotted line),
 $\omega_0=0.625\Gamma$ (dashed line) for a fixed AC amplitude 
 $V_{AC}=0.5\Gamma$.}
\label{fig:3}
\end{figure}

\begin{figure}[h]
\centerline{
\psfig{figure=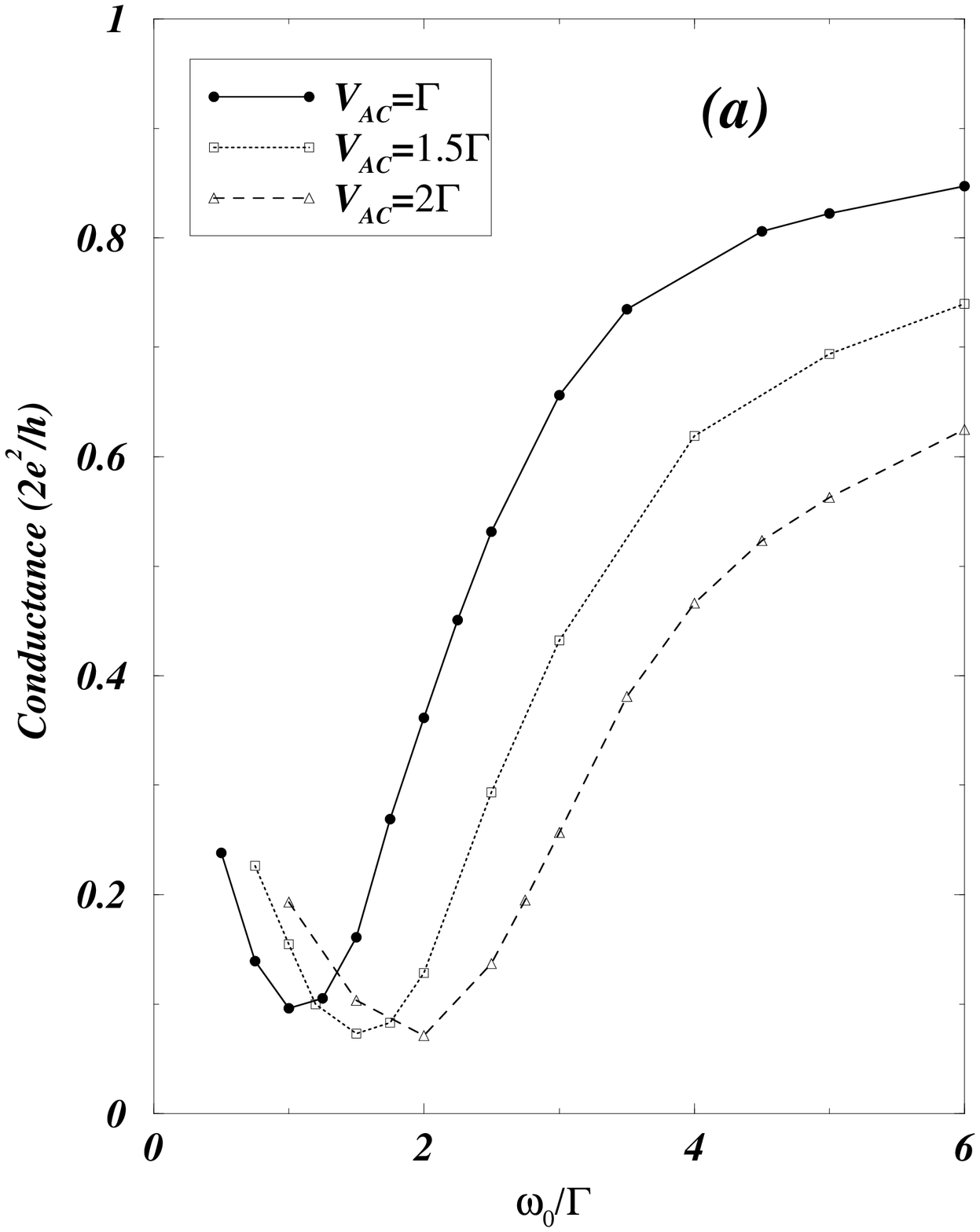, height=8.0cm, width=8.0 cm}
\psfig{figure=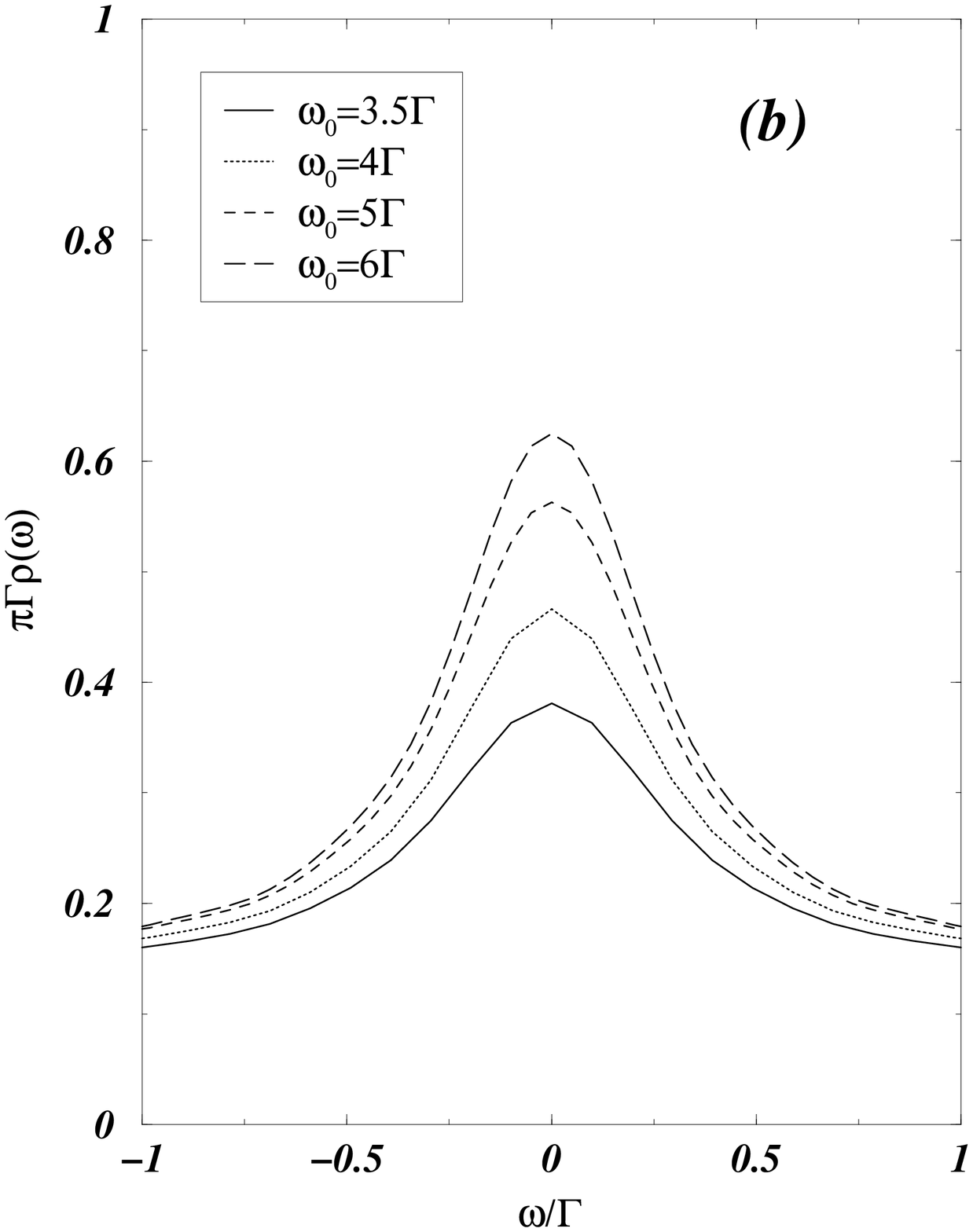, height=8.0cm, width=8.0 cm}
}
\caption{(a) Conductance as a function of $\omega_0$ in the Kondo regime, $U=2.5\pi\Gamma$ and $T=0.05\Gamma$. The solid line shows the case of $V_{AC}=\Gamma$,
the dotted line corresponds to the case of $V_{AC}=1.5\Gamma$
and finally dashed line shows the strongest AC intensity $V_{AC}=2\Gamma$.\\
(b) Time-averaged DOS for the strongest field amplitude $V_{AC}=2\Gamma$ for different large AC frequencies. The solid line 
shows the case of $\omega_0=3.5\Gamma$ ($\beta=0.57$) where the Kondo peak has 
the lower height, the dotted line shows the case $\omega_0=4\Gamma$ 
($\beta=0.5$), the dashed line corresponds to $\omega_0=5\Gamma$ 
($\beta=0.22$) and the long-dashed line depicts the case of $\omega_0=6\Gamma$ ($\beta=0.16$).} 
\label{fig:4}
\end{figure}

\begin{figure}[t]
\centerline{
\psfig{figure=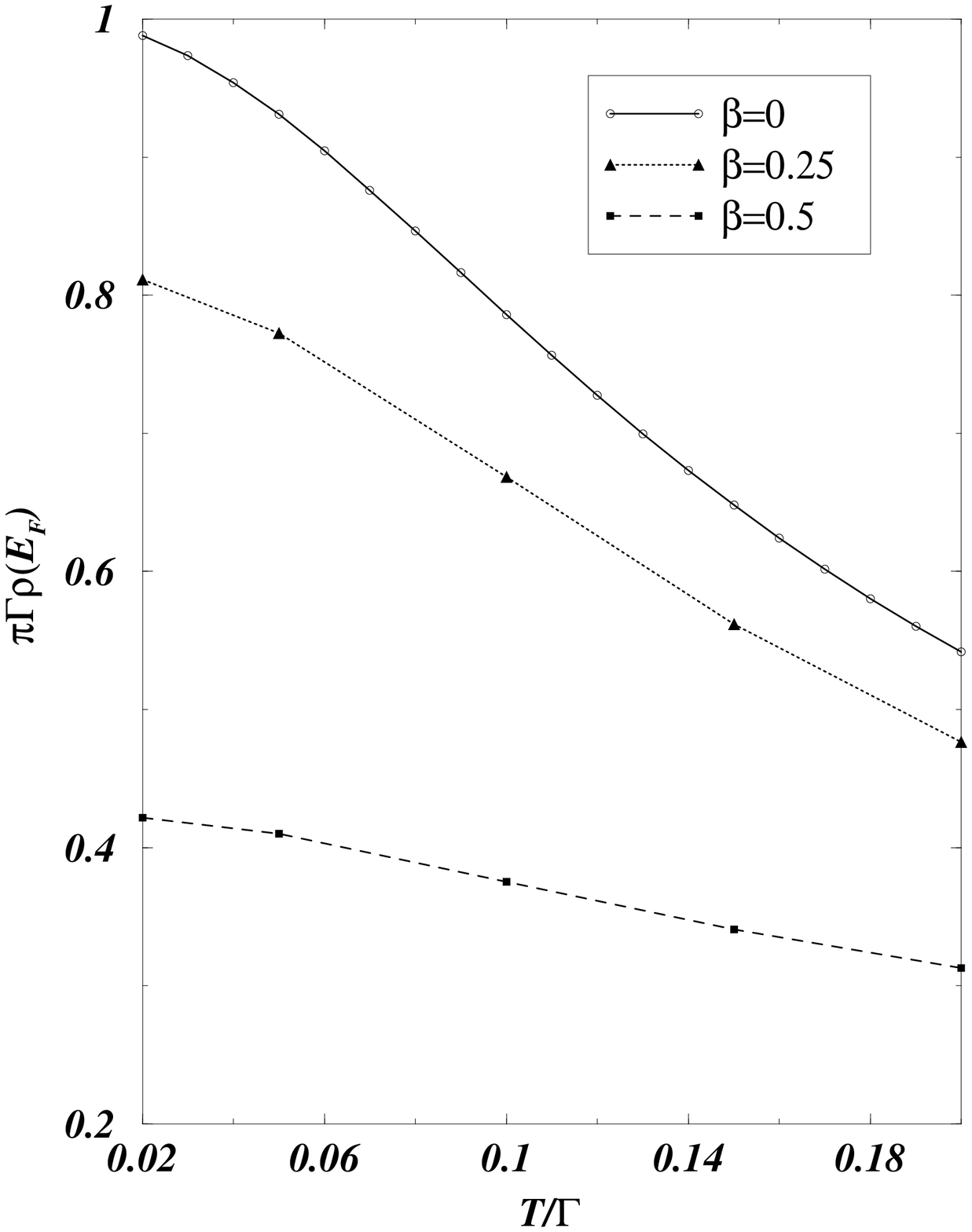, height=8.0cm, width=8.0cm}
}
\caption{Time-averaged DOS at $E_F$ for $U=2.5\pi\Gamma$ as a function of temperature. The solid line shows the case in the absence of the AC potential. The 
dotted line corresponds to a case of a low AC intensity  
 $V_{AC}=0.25\Gamma\approx 2T_K$ ($\beta=0.25$) and the dashed line
 shows a case with higher AC intensity $V_{AC}=0.5\Gamma\approx 
4T_K$ ($\beta=0.5$). Both of them correspond to an AC frequency 
$\omega_0=\Gamma>>T_K$.}
\label{fig:5}
\end{figure}

We solve numerically the set of linear equations (\ref{discretized}) for a QD in the Kondo regime 
(symmetric case $\epsilon_0=-U/2$ with symmetric couplings $\Gamma_L=\Gamma_R=\Gamma$ and U=2.5$\pi \Gamma$) 
and a temperature T=$0.05\Gamma$ \cite{Fer} for different AC frequencies and AC intensities. From the solution of Eq. (\ref{discretized}) for the retarded Green's function we obtain the
time-averaged density of states (Eq. 21) and the linear conductance (Eq. 22). \\ 
The main effect of the AC potential consists in a reduction of the time-averaged DOS at $E_F$. 
This reduction can be interpreted as decoherence induced by AC excitations, either by real photon-assisted induced excitations at large AC frequencies \cite{Win2} or virtual spin-flip cotunneling processes at small 
AC frequencies \cite{Gla2}. 
These processes introduce a quenching of the Kondo peak causing a deviation of the linear conductance 
from the unitary limit. 
This suppression induced by the AC potential depends on two competing mechanisms regulated 
by the AC frequency and by the absorption rate of photons. 
If the AC frequency is small ($\omega_0\leq T_K$), a negligible reduction will take place even for a large absorption rate of photons 
(the probability of absorption depends on Bessel functions with argument $\beta=V_{AC}/\omega_0$, as shown in appendix B).
On the other hand, for large AC frequencies the mechanism of suppression (AC excitations) 
becomes ineffective as $\beta\rightarrow 0$.\\
To analyze in more detail the behavior of the time-averaged DOS as a function of the AC parameters, we plot
in Fig. \ref{fig:1} the time-averaged DOS for three different AC frequencies, 
$\omega_0=\Gamma /4\approx 2T_K$ (Fig. \ref{fig:1}a), $\omega_0=\Gamma/2 \approx 4T_K$ (Fig. \ref{fig:1}b) and 
$\omega_0=\Gamma\approx 8T_K$ (Fig. \ref{fig:1}c). For each AC frequency we consider different 
AC intensities corresponding to $\beta$= 0, 0.25, 0.5 and 0.75.
Fig. \ref{fig:1}a corresponds to $\omega_0=\Gamma /4\approx 2 T_K$ and different AC intensities 
$V_{AC}\approx T_K/2$ ($\beta=0.25$),  $V_{AC}\approx T_K $ ($\beta=0.5$) and $V_{AC}\approx 3T_K/2$ ($\beta=0.75$). 
In all these situations the Kondo
peak is slightly reduced, i. e. the dynamics of the correlated
collective state is practically not affected by the AC potential since both, the AC frequency and the AC intensities 
are of the order of $T_K$.  
Furthermore, there is
no evidence of replicas of the Kondo peak in the time-averaged DOS.  
By doubling $\omega_0$ (Fig. \ref{fig:1}b), the Kondo peak
undergoes a stronger reduction as the AC intensity grows.
In the present case, the two external energy scales, $\omega_0$ and $V_{AC}$, are larger than $T_K$
yielding to a stronger modification of the time-averaged DOS at $E_F$. 
However, a total suppression of the Kondo resonance is only reached for our largest AC frequency 
(Fig. \ref{fig:1}c, $\omega_0=\Gamma$). 
Even for a small value of $\beta=0.25$, the Kondo resonance is reduced remarkably. 
For the highest AC amplitude ($\beta=0.75$) the Kondo resonance has been destroyed completely. 
Another important effect induced by the AC signal in the time-averaged 
DOS is that the Kondo peak develops satellites. 
These satellites become apparent in Figs. \ref{fig:1}b and \ref{fig:1}c. In these cases both $\omega_0$ and $V_{AC}$ are much larger than
the relevant energy scale of this problem given by $T_K$. 
In Fig. \ref{fig:1}c  even the second satellites are resolved at
$\beta=0.5$ \cite{Sat}.
In a mean field model, one should expect an appreciable 
increase of the satellites of the main peaks in the time-averaged DOS as $\beta$ increases \cite{Tien1,Tien2}. 
However, in the present case, 
the satellites grow very slowly with $\beta$ due to the two competing mechanisms previously discussed.
It is important to note that, the mean field resonances located at 
$\epsilon_0$ and $\epsilon_0+\frac{U}{2}$ do not display satellites in the
time-averaged DOS, since $\omega_0$ is smaller that their widths which are of the 
order of $2\Gamma$.
\\
To illustrate the previous discussion, we plot in Fig. \ref{fig:2} the conductance as a 
function of $V_{AC}$ for four different 
AC frequencies. The height of the Kondo peak falls off as the applied AC intensity increases. 
Moreover, the suppression of the time-averaged DOS at $E_F$ occurs in a very similar way for the low ($\omega_0=\Gamma /4$) and intermediate AC frequencies ($\omega_0=\Gamma /2$, $\omega_0=\Gamma$). 
This decrease is slower for the largest AC frequency ($\omega_0=1.75\Gamma$), 
since for each AC amplitude the absorption rate of photons ,which depends on $\beta$, 
is the smallest for the largest AC frequency.  \\
More interesting is the analysis of the conductance as a function
of $\omega_0$, at a fixed AC intensity (in Fig. \ref{fig:3}a for $V_{AC}<\Gamma$ and in Fig. \ref{fig:4}a 
for $V_{AC}\geq\Gamma$). 
In both cases we find that there are two regimes separated by a threshold frequency, $\omega_t$, where the 
conductance has a minimum. 
The presence of this minimum can be understood, again, by the competition of two opposite effects.
In the adiabatic limit, where $\omega_0\rightarrow 0$,
the effect of the applied AC potential has a negligible effect on the dynamics of the
correlated collective state. 
In this case, the quenching of the Kondo peak is no longer effective and, therefore,
the AC only induces a small decoherence in the correlated state \cite{Gla2}.
In the very high AC frequency limit, where
$\omega_0\rightarrow\infty$ (in this limit $\beta\rightarrow 0$), the unpaired electron has a negligible absorption 
probability. 
Thus, it is obvious from the previous discussion that the conductance versus $\omega_0$ should display a minimal value. 
For low AC intensities (Fig. \ref{fig:3}a), the threshold frequency depends on the AC amplitude : for $V_{AC}=0.25\Gamma$ $\omega_t\approx 0.5\Gamma$ and for $V_{AC}=0.5\Gamma$ 
$\omega_t\approx 0.625\Gamma$. In both cases, for AC frequencies lower than $\omega_t$ the conductance decreases 
as $\omega_0$ increases whereas this tendency is reversed for AC frequencies larger than $\omega_t$. 
Note that for $\omega_0<\omega_t$ the slowest drop of conductance corresponds 
to the smallest AC amplitude. For $\omega_0>\omega_t$ the conductance increases with increasing $\omega_0$. Now, the conductance recovers its undriven limit for smaller AC frequencies in the case of the smallest AC intensity as one should expect. 
The position of the thresholds frequency depends on the competition of the two effects previously discussed. 
This competition is illustrated in Fig. \ref{fig:3}b,  where we plot the time-averaged DOS at $E_F$ for three AC frequencies 
$\omega_0\leq \omega_t$. 
By increasing the AC frequency from $\omega_0=0.375\Gamma$ up to $\omega_0=0.5\Gamma$ the Kondo resonance is 
suppressed more effectively than by raising the AC frequency from $\omega_0=0.5\Gamma$ up to $\omega_0=0.625\Gamma$ 
where the reduction of the Kondo peak is almost negligible. 
Fig. \ref{fig:4}a depicts the conductance vs. $\omega_0$ for $V_{AC}\geq\Gamma$. 
Remarkably, if the threshold frequency is normalized to the AC amplitude we find that $\omega_t$ decreases extremely slowly for the three AC intensities ($\omega_0/V_{AC}\simeq 1)$. 
Here, again the suppression of the conductance is slower for the lowest AC amplitude. 
Furthermore, in the region where the conductance is an increasing function of the AC frequency, 
this increase is slower for the highest AC intensity. 
The previous behavior is illustrated
in Fig. \ref{fig:4}b where we plot the time-averaged DOS for several AC frequencies at the strongest AC intensity $V_{AC}=2\Gamma$.\\ 
Let us now compare our results with the ones obtained with different schemes \cite{Gla2,Win2}:  
Kaminski {\it et al.} \cite {Gla2} predict no suppression of the Kondo peak for the 
symmetric case ($\epsilon _0=-U/2$) in clear disagreement with both our theory and experimental results\cite{Sil}. In asymmetric cases, they predict a monotonous decrease of
the height of the Kondo peak as a function of $\omega_0$ (in the low AC frequency regime at fixed $V_{AC}$).
Their results (for $\epsilon _0 \neq -U/2$) are similar to our results (for $\epsilon _0=-U/2$) for AC frequencies below
$\omega_0\approx \Gamma/2$ and for low AC intensities (solid line in Fig. \ref{fig:3}a) 
where the Kondo peak is not strongly suppressed. The rest of the cases discussed here are away from the range of 
validity of Ref. \cite {Gla2}. \\
Using a NCA, Nordlander {\it et al.} \cite{Win2} 
find a nonmonotonous behavior of the 
conductance vs. $\omega_0$ and the existence of a minimum, in qualitative agreement with us. 
Caution is needed, however, in comparing both works due to the different approximations involved and the 
different regimes of validity
for both calculations.
Whereas our perturbative calculation is valid for the symmetric case, i. e. particle-hole symmetry, and low temperatures, 
the NCA considers the limit $U\rightarrow\infty$ (i. e. strongly asymmetric case) and temperatures of the order and higher than $T_{K}$.
In that sense, both calculations should be regarded as complementary.\\
In Fig. \ref{fig:5} we plot the time-averaged DOS at $E_F$ as a function of $T$ in the absence of the AC potential (solid line) and in the presence of the AC potential
for $\omega_0=\Gamma$, $V_{AC}=0.25\Gamma$ (dashed line) and $V_{AC}=0.5\Gamma$  (dashed line).
In the absence of the AC potential, we get the well known strong reduction of 
the DOS at $E_F$ when the temperature increases, as expected \cite{Hew}. 
For the lowest AC intensity case (dotted line) a strong temperature dependence 
is observed similarly to the case without AC potential (solid line).
However, such a dependence changes for the strongest AC intensity (dashed line).
In this case, the time-averaged DOS
at $E_F$ has been already so strongly reduced
at very low temperatures ($T=0.02\Gamma$) that an increase of temperature 
produces only a small further reduction of the linear conductance. This result can be expected for a system where the Kondo effect is already very
weak at low temperatures .
This behavior
of the Kondo peak height as a function of temperature has been confirmed experimentally (Fig. 8 in Ref \cite{Sil}).

\section{Summary}
In conclusion, new features are found in the finite temperature transport ($T<<T_K$) 
properties of QD's in the Kondo regime as an 
oscillatory gate voltage is applied. 
By solving the time dependent
Dyson equation we obtain the QD retarded Green's function and the time-averaged
density of states within the framework of the Fermi liquid theory.
The interaction self-energy in our model is calculated, in the context of the Keldysh non-equilibrium technique, 
by perturbation theory
up to second order in the on-site interaction and the effect of the AC
potential is taken into account exactly for {\it all
ranges of AC frequencies and AC intensities of the AC potential}.
The Kondo resonance is modified by the external AC voltage in a different way depending on the
range of AC frequencies studied.
We find two different AC frequency ranges of opposite behavior, separated by a threshold
frequency $\omega_{t}$ where the linear conductance is minimum. 
$\omega_{t}$ depends on the AC intensity
and moves to higher values as $V_{AC}$ increases. 
At small AC frequencies, and fixed AC intensity, the Kondo peak decreases
as $\omega_{0}$ grows. Once the threshold frequency
is reached the opposite behavior is found and the Kondo peak increases 
as $\beta=\frac{V_{AC}}{\omega_0}\rightarrow 0$. 
Our results qualitatively agree with previous theories and complete the range of
parameters (arbitrary AC frequencies and AC intensities at finite on-site interaction) 
and the regime of temperatures (temperatures below $T_{K}$).
We also analyze the behavior of the Kondo peak as a function of
temperature at fixed AC frequency. For large AC intensities, we obtain a very 
small decrease of the Kondo peak as temperature increases, while for small 
AC intensities, a much larger quenching of the peak is obtained.

\section{Acknowledgements}
Work supported in part by the MEC of Spain under contracts
PB96-0875, PB96-0085 and grant PF 98-07497938, the CAM under contract No. 07N/0026/1998; 
by the NSF grant No. DMR 97-08499
and by the EU via contract FMRX-CT98-0180. 
We gratefully acknowledge discussions with David Sanchez and Silvano De Franceschi.

\appendix
\section{QD Green's functions in the absence of AC potential}
In this appendix, the QD Green's functions in the absence of time modulation 
are obtained. First of all, the QD Green's function coupled to the leads is 
calculated, and afterwards the interaction in the Hartree approximation is included.
A very simple calculation yields the exact lesser and greater Green's functions 
for $H_{leads}$,
\begin{eqnarray}\label{glead}
&& g_{k}^{<,(0)}(t-t')  \equiv
i\langle{c}_{k}^{\dagger}(t'){c}_{k}(t)\rangle=
i f(\epsilon_k) e^{-i\epsilon_{k}(t-t')},
\nonumber\\
&& g_{k}^{>,(0)}(t-t')  \equiv
-i\langle {c}_{k}(t) {c}_{k}^{\dagger}(t')\rangle= -i(1-f(\epsilon_k))e^{-i\epsilon_{k}(t-t')},
\end{eqnarray}
where $f(\epsilon_k)$ is the Fermi-Dirac distribution function.
The lesser and greater hopping self-energies including hopping between the 
QD and the leads are written in terms  
of the previous Green's functions as\cite{Jau}, 
\begin{eqnarray}\label{hopp_self<}
\small{\Sigma}^<_{sd}(t_1-t_2) = \sum_{k\in L,R}
V^*_k g^<_k(t_1-t_2)V_k\;,=
 i\sum_{L,R} \int {d\epsilon\over 2\pi}
e^{-i\epsilon(t_1-t_2)}\times f_{L/R}(\epsilon)\Gamma^{L/R}(\epsilon)\;,
\end{eqnarray}
\begin{eqnarray}\label{hopp_self>}
\Sigma^>_{sd}(t_1-t_2) = \sum_{k\in L,R}
V^*_{k}g^>_{k}(t_1-t_2)V_{k}\;,= -i\sum_{L,R} \int {d\epsilon\over 2\pi}
{e}^{-i\epsilon(t_1-t_2)}\times (1-f_{L/R}(\epsilon))\Gamma^{L/R}(\epsilon)\;.
\end{eqnarray}
where $\Gamma^L(\epsilon)=2\pi\sum_{k\in L}|V_{k}|^2\delta(\epsilon-\epsilon_k)$ and a similar expression is obtained for 
$\Gamma^R(\epsilon)$.
The retarded and advanced self-energies fulfill the relations 
\begin{eqnarray}
&&\small{\Sigma}^r_{sd}(t_1-t_2)=\theta(t_1-t_2)(\small{\Sigma}^>_{sd}(t_1-t_2)-{\Sigma}^<_{sd}(t_1-t_2)),
\nonumber\\
&&\small{\Sigma}^a_{sd}(t_1-t_2)=\theta(t_2-t_1)(\small{\Sigma}^<_{sd}(t_1-t_2)-{\Sigma}^>_{sd}(t_1-t_2)).
\end{eqnarray}
The Dyson equation for the retarded and advanced QD Green's functions is
\begin{eqnarray}\label{Gret}
g_{d,\sigma}^{r,a} && (t-t') \,  = g^{r,a,(0)}_{d,\sigma}(t-t') +
\int dt_1 \int dt_2 g^{r,a,(0)}_{d,\sigma}(t-t_1)
\Sigma_{sd}^{r,a}(t_1-t_2)g_{d,\sigma}^{r,a}(t_2-t')\;.
\end{eqnarray}
Here $g^{r,a,(0)}_{d,\sigma}(t-t')$ are the QD Green's functions for 
$H_{QD}$ (Eq. (6)) without the on-site repulsion term,
\begin{eqnarray}
g^{r,a,(0)}_{d,\sigma}(t-t')=\mp i\theta(\pm t \mp t')
e^{-i\epsilon_{0,\sigma}(t-t')}.
\end{eqnarray}
In the absence of time dependence, it is advantageous to Fourier transform, getting 
\begin{eqnarray}\label{time_ind}
{g}^{r,a}_{d,\sigma}(\epsilon) =[({
g}^{r,a,(0)}_{d,\sigma})^{-1}-{\Sigma}^{r,a}_{sd}(\epsilon)]^{-1},
\end{eqnarray}
where
\begin{equation}\label{sigma}
\Sigma^{r,a}_{sd}(\epsilon) = \sum_{k\in L,R}
\frac{|V_{k}|^2}{\epsilon-\epsilon_k\pm i\eta}=\;\sum_{\alpha\in L,R}\Lambda_\alpha(\epsilon) \mp {i\over 2}\Gamma_\alpha(\epsilon)\;.
\end{equation}
In the WB limit approximation the previous self-energies become 
$\Sigma_{sd}^{r,a}(\epsilon) = \mp \sum_{\alpha\in L,R}{i\over 2}\Gamma_\alpha$. 
By inserting these self-energies in Eq. (\ref{time_ind}), the 
retarded and advanced QD Green's functions coupled to the leads are 
\begin{equation}\label{gracoplada}
g^{r,a}_{d,\sigma}(\epsilon) = \frac{1}{\epsilon - \epsilon_{0,\sigma} \pm {i \over 2} \sum_{\alpha\in L,R} \Gamma_\alpha}.
\end{equation}
One can Fourier transform back to get the time dependence of the retarded and 
advanced QD Green's functions 
\begin{eqnarray}
g^{r,a}_{d,\sigma}(t,t')= \mp i\theta(\pm t \mp t')e^{-i\int_{t'}^{t}dt_1(\epsilon_{0,\sigma}\mp i\sum_{\alpha\in L,R}\Gamma_\alpha/2)}\;.
\end{eqnarray}
Now the on-site interaction self-energy in the Hartree approximation 
is calculated via perturbation theory up to first order in U: 
\begin{eqnarray}
\Sigma^{r,a(1)}_{d,\sigma}=U n_{d,\bar{\sigma}},
\end{eqnarray}
where $n_{d,\bar\sigma}=\langle d_{\bar\sigma}^\dagger(t) 
d_{\bar\sigma}(t)\rangle$. By using the Dyson equation it is straightforward to get
\begin{eqnarray}\label{Hartree}
{\mathbf {g}}^{r,a}_{d,\sigma}(t,t')=  \mp i\theta(\pm t \mp t')e^{-i\int_{t'}^{t}dt_1(\epsilon_{0,\sigma}+Un_{d,\bar{\sigma}}\mp i\sum_{\alpha\in L,R}\Gamma_\alpha/2)}\;.
\end{eqnarray} 

\section{QD Green's functions in the presence of AC potential}
In this appendix the analytical expressions for the lesser and greater 
QD Green's functions in the presence of an AC potential, coupled to the 
leads in the Hartree approximation are derived. As we said in Sec. II, one 
needs to obtain these propagators in order to have an expression of the 
interaction self-energy which is given by Eq. (\ref{int_self}). However, 
the lesser and greater QD Green's function are given as a function of the 
retarded and advanced self-energies (see Eq. (\ref{less_grea_GF})). 
Therefore, the first step in this derivation is to compute the retarded and 
advanced QD Green's functions. 
\subsection{retarded and advanced QD Green's Functions}
One way to obtain the retarded QD Green's function (similar analysis can 
be done for the advanced QD Green's function) is by means of perturbation 
theory in the AC signal up to infinite order. For this purpose, one 
takes the QD Green's function (\ref{Hartree}) as the bare QD Green's function. 
Using $\Sigma^r_{AC}(t)=-iV_{AC}cos \omega_0 t$ $\;$
($\Sigma_{AC}^r(t)=\Sigma_{AC}^C(t)-\Sigma_{AC}^<(t)$, where $\Sigma_{AC}^<(t)
=0$), as the retarded self-energy, 
the Dyson equation for the retarded QD Green's function reads 
\begin{eqnarray} \label{retarded_GF}
&&G^r_{d,\sigma}(t,t^\prime)= {\mathbf {g}}^{r}_{d,\sigma}(t-t^\prime)-i\int_{-\infty}^{\infty} d\tau {\mathbf {g}}^{r}_{d,\sigma}(t-\tau)\; V_{AC}cos \omega_0 \tau\; {\mathbf {g}}^r_{d,\sigma}(\tau-t^\prime)- \nonumber \\
&& \frac{1}{2} \int_{-\infty}^{\infty} d\tau \int_{-\infty}^{\infty} d\tau_1\; {\mathbf {g}}^{r}_{d,\sigma}(t-\tau)\; V_{AC}cos \omega_0 \tau\; {\mathbf {g}}^{r}_{d,\sigma}(\tau-\tau_1)\; V_{AC}cos \omega_0 \tau_1\; {\mathbf {g}}^{r}_{d,\sigma}(\tau_1-t^\prime)\;+ ...
\end{eqnarray}
Eq. (\ref{retarded_GF}) can be written as
\begin{eqnarray}\label{retarded_sum}
G^r_{d,\sigma}(t,t') & = &  {\mathbf {g}}^r_{d,\sigma}(t-t^\prime)\;\Bigl[1 -i  \int_{t}^{t^\prime} V_{AC}cos \omega_0 \tau  d\tau - 
\nonumber\\
&&\frac{1}{2} \int_t^{t'} V_{AC}cos \omega_0 \tau d\tau \int_t^{t'} V_{AC}cos \omega_0 \tau_1  d\tau_1+ ...\Bigr].
\end{eqnarray}
Performing the summation, one obtains the time dependent retarded 
QD Green's function:
\begin{eqnarray}\label{retarded_exact}
G^r_{d,\sigma}(t,t^\prime)&=&e^{-i\frac{V_{AC}}{\omega_0}sin\omega_0 t} e^{i\frac{V_{AC}}{\omega_0}sin\omega_0 t'}\; {\mathbf {g}}^{r}_{d,\sigma}(t-t^\prime)=
\nonumber\\
&&\sum_{p=-\infty}^{p=\infty}\sum_{m=-\infty}^{m=\infty}J_m(\beta)J_p(\beta)e^{-i p\omega_0 t}e^{i m\omega_0 t'}{\mathbf {g}}^{r}_{d,\sigma}(t-t^\prime).
\end{eqnarray} 
where $J_m(\beta)$ is the Bessel function of order m and argument $\beta=\frac{V_{AC}}{\omega_0}$.
A similar procedure yields the time dependent advanced QD Green's function
\begin{eqnarray}\label{advanced_exact}
G^a_{d,\sigma}(t,t^\prime)&=&e^{-i\frac{V_{AC}}{\omega_0}sin\omega_0 t} e^{i\frac{V_{AC}}{\omega_0}sin\omega_0 t'}\; {\mathbf {g}}^{a}_{d,\sigma}(t-t^\prime)=
\nonumber\\
&&\sum_{p=-\infty}^{p=\infty}\sum_{m=-\infty}^{m=\infty}J_m(\beta)J_p(\beta)
e^{-i p\omega_0 t}e^{i m\omega_0 t'}{\mathbf {g}}^{a}_{d,\sigma}(t-t^\prime).
\end{eqnarray} 
\subsection{greater and lesser QD Green's functions}
The greater and lesser QD Green's functions are calculated by substituting 
the retarded and advanced QD Green's functions (Eqs. (\ref{retarded_exact}), 
(\ref{advanced_exact})) in Eq. (\ref{less_grea_GF}): 
\begin{eqnarray}\label{lesser_GF}
G^<_{d,\sigma}(t,t')&=&e^{-i\frac{V_{AC}}{\omega_0}sin\omega_0 t} e^{i\frac{V_{AC}}{\omega_0}sin\omega_0 t'}\sum_{p=-\infty}^{p=\infty}\sum_{m=-\infty}^{m=\infty}J_m(\beta)J_p(\beta)\sum_{L,R}
\nonumber\\
&&\int_{-\infty}^{\infty}\frac{d\epsilon}{2\pi}\frac{e^{i p\omega_0 t}e^{-i m\omega_0 t'}
e^{-i\epsilon(t-t')}f_{L/R}(\epsilon)\Gamma^{L/R}(\epsilon)}{(\epsilon-p\omega_0-\bar\epsilon_{0,\sigma}
+{i \over 2} \sum_{\alpha\in L,R} \Gamma_\alpha)(\epsilon-m\omega_0-\bar\epsilon_{0,\sigma}
-{i \over 2} \sum_{\alpha\in L,R} \Gamma_\alpha)},
\end{eqnarray}
\begin{eqnarray}\label{greater_GF}
G^>_{d,\sigma}(t,t')&=&e^{-i\frac{V_{AC}}{\omega_0}sin\omega_0 t} e^{i\frac{V_{AC}}{\omega_0}sin\omega_0 t'}\sum_{p=-\infty}^{p=\infty}\sum_{m=-\infty}^{m=\infty}J_m(\beta)J_p(\beta)\sum_{L,R}
\nonumber\\
&&\int_{-\infty}^{\infty}\frac{d\epsilon}{2\pi}\frac{e^{i p\omega_0 t}e^{-i m\omega_0 t'}
e^{-i\epsilon(t-t')}(1-f_{L/R}(\epsilon))\Gamma^{L/R}}{(\epsilon-p\omega_0-\bar\epsilon_{0,\sigma}
+{i \over 2} \sum_{\alpha\in L,R} \Gamma_\alpha)(\epsilon-m\omega_0-\bar\epsilon_{0,\sigma}
-{i \over 2} \sum_{\alpha\in L,R} \Gamma_\alpha)},
\end{eqnarray}
\begin{eqnarray}\label{lesser_GF_inv}
G^<_{d,\sigma}(t',t)&=&e^{i\frac{V_{AC}}{\omega_0}sin\omega_0 t} e^{-i\frac{V_{AC}}{\omega_0}sin\omega_0 t'}\sum_{p=-\infty}^{p=\infty}\sum_{m=-\infty}^{m=\infty}J_m(\beta)J_p(\beta)\sum_{L,R}
\nonumber\\
&&\int_{-\infty}^{\infty}\frac{d\epsilon}{2\pi}\frac{e^{-i p\omega_0 t}e^{i m\omega_0 t'}
e^{-i\epsilon(t'-t)}f_{L/R}(\epsilon)\Gamma^{L/R}}{(\epsilon-p\omega_0-\bar\epsilon_{0,\sigma}
+{i \over 2} \sum_{\alpha\in L,R} \Gamma_\alpha)(\epsilon-m\omega_0-\bar\epsilon_{0,\sigma}
-{i \over 2} \sum_{\alpha\in L,R} \Gamma_\alpha)},
\end{eqnarray}
\begin{eqnarray}\label{greater_GF_inv}
G^>_{d,\sigma}(t',t)&=&e^{i\frac{V_{AC}}{\omega_0}sin\omega_0 t} e^{-i\frac{V_{AC}}{\omega_0}sin\omega_0 t'}\sum_{p=-\infty}^{p=\infty}\sum_{m=-\infty}^{m=\infty}J_m(\beta)J_p(\beta)\sum_{L,R}
\nonumber\\
&&\int_{-\infty}^{\infty}\frac{d\epsilon}{2\pi}\frac{e^{-i p\omega_0 t}e^{i m\omega_0 t'}
e^{-i\epsilon(t'-t)}(1-f_{L/R}(\epsilon))\Gamma^{L/R})}{(\epsilon-p\omega_0-\bar\epsilon_{0,\sigma}
+{i \over 2} \sum_{\alpha\in L,R} \Gamma_\alpha))(\epsilon-m\omega_0-\bar\epsilon_{0,\sigma}
-{i \over 2} \sum_{\alpha\in L,R} \Gamma_\alpha))}.
\end{eqnarray}
Once again we have taken $\bar\epsilon_{0,\sigma}=
\epsilon_{0,\sigma}+U n_{d,\bar\sigma}$ being $n_{d,\bar\sigma}$ the QD 
occupation. 
In the case of zero DC bias, the right and left Fermi-Dirac distribution functions are 
$f_{L,R}(\epsilon)=\frac{1}{e^{(\epsilon-E_F)/T}+1}$
where $T$ is the temperature. 
The analytical expressions for the lesser 
and greater QD Green's function are obtained by integrating in the complex 
plane where the Fermi-Dirac distribution function can be written as a
difference of two digamma functions which have poles
in the lower and upper complex plane:
\begin{eqnarray}\label{digammas}
f(z)-{1 \over 2}=-{1 \over {2\pi i}}\Bigl[\psi ({1 \over 2}+ {iz \over {2\pi T}})
-\psi({1 \over 2}-{iz \over {2\pi T}})\Bigr]=
\sum_{k=0}^{k=\infty} {1 \over {{1 \over 2}+k-{iz \over {2\pi T}}}}
-{1 \over {{1 \over 2}+k+{iz \over {2\pi T}}}}.
\end{eqnarray}
The complex integrals are performed with the restriction $t-t'\geq 0$ because we use them
to evaluate a retarded quantity (the retarded self-energy in Eq. (12)).\\ 
In order to abbreviate the notation, we define the following variables 
\begin{eqnarray}\label{variables}
&&a_p=-(\bar\epsilon_{0,\sigma}+p\omega_0-{i \over 2} \sum_{\alpha\in L,R} \Gamma_\alpha),\; b_m=-(\bar\epsilon_{0,\sigma}+m\omega_0+{i \over 2} \sum_{\alpha\in L,R} \Gamma_\alpha),
\nonumber
\\
&&\alpha_p=\frac{1}{2}+\frac{i}{2\pi T}a_p, \; \gamma_m=\frac{1}{2}+\frac{i}{2\pi T}b_m,\; \zeta=e^{-2\pi T(t-t')},
\nonumber\\
&&\bar{u}_p=\frac{1}{2}(1-tanh(\frac{1}{2T} a_p)),\; \bar{v}_m=\frac{1}{2}(1-tanh(\frac{1}{2T} b_m)).
\end{eqnarray}\label{analitical_lesser}
With this notation, the analytic expressions for the lesser and greater QD Green's functions are:
\begin{eqnarray}
&&G^>(t,t')=e^{-i\frac{V_{AC}}{\omega_0}sin\omega_0 t} e^{i\frac{V_{AC}}{\omega_0}sin\omega_0 t'}\sum_{p=-\infty}^{p=\infty}\sum_{m=-\infty}^{m=\infty}J_m(\beta)J_p(\beta)\frac{e^{i p\omega_0 t}e^{-i m\omega_0 t'}}{(p-m)\omega_0-i\sum_{\alpha\in L,R} \Gamma_\alpha}
\nonumber\\
&&\times(\frac{F_1(\alpha_p,1,\alpha_p+1,\zeta)}{\alpha_p}-\frac{F_1(\gamma_m,1,\gamma_m+1,\zeta)}{\gamma_m}-2\pi i\; (\bar{u}_p-1)\; e^{i a_p(t-t')}),
\end{eqnarray}
\begin{eqnarray}\label{analitical_greater}
&&G^<(t,t')=e^{-i\frac{V_{AC}}{\omega_0}sin\omega_0 t} e^{i\frac{V_{AC}}{\omega_0}sin\omega_0 t'}\sum_{p=-\infty}^{p=\infty}\sum_{m=-\infty}^{m=\infty}J_m(\beta)J_p(\beta)\frac{e^{i p\omega_0 t}e^{-i m\omega_0 t'}}{(p-m)\omega_0-i\sum_{\alpha\in L,R} \Gamma_\alpha}
\nonumber\\
&&\times(\frac{F_1(\alpha_p^*,1,\alpha_p^*+1,\zeta)}{\alpha_p^*}-\frac{F_1(\gamma_m^*,1,\gamma_m^*+1,\zeta)}{\gamma_m^*}-2\pi i \;\bar{v}_m \;e^{-i b_m(t-t')}),
\end{eqnarray}
\begin{eqnarray}\label{analitical_lesser_inv}
&&G^>(t',t)=e^{i\frac{V_{AC}}{\omega_0}sin\omega_0 t} e^{-i\frac{V_{AC}}{\omega_0}sin\omega_0 t'}\sum_{p=-\infty}^{p=\infty}\sum_{m=-\infty}^{m=\infty}J_m(\beta)J_p(\beta)\frac{e^{-i p\omega_0 t}e^{i m\omega_0 t'}}{(p-m)\omega_0-i\sum_{\alpha\in L,R} \Gamma_\alpha}
\nonumber\\
&&\times(\frac{F_1(\alpha_p,1,\alpha_p+1,\zeta)}{\alpha_p}-\frac{F_1(\gamma_m,1,\gamma_m+1,\zeta)}{\gamma_m}-2\pi i \; \bar{u}_p \;e^{i a_p(t-t')}),
\end{eqnarray}
\begin{eqnarray}\label{analitical_greater_inv}
&&G^<(t',t)=-e^{i\frac{V_{AC}}{\omega_0}sin\omega_0 t} e^{-i\frac{V_{AC}}{\omega_0}sin\omega_0 t'}\sum_{p=-\infty}^{p=\infty}\sum_{m=-\infty}^{m=\infty}J_m(\beta)J_p(\beta)\frac{e^{-i p\omega_0 t}e^{i m\omega_0 t'}}{(p-m)\omega_0-i\sum_{\alpha\in L,R} \Gamma_\alpha}
\nonumber\\
&&\times(\frac{F_1(\alpha_p^*,1,\alpha_p^*+1,\zeta)}{\alpha_p^*}-\frac{F_1(\gamma_m^*,1,\gamma_m^*+1,\zeta)}{\gamma_m^*}-2\pi i\;(\bar{v}_m-1)\; e^{-i b_m(t-t')}).
\end{eqnarray}
The functions $F_1$ are hypergeometric functions. Here $\gamma^*$ and $\alpha^*$ are complex conjugates of  
$\gamma$ and $\alpha$ respectively.
Once the greater and the lesser QD Green's functions are obtained, 
the retarded QD interaction self-energy is given by: 
\begin{eqnarray}\label{int_selfret}
\Sigma^{r,(2)}_{d,\sigma}(t,t')=-iU^2\theta(t-t')[G^<_{d,\sigma}(t,t')G^>_{d,\bar{\sigma}}(t',t)G^<_{d,\bar\sigma}(t,t')
-G^>_{d,\sigma}(t,t')G^<_{d,\bar{\sigma}}(t',t)G^>_{d,\bar\sigma}(t,t')].
\end{eqnarray}
Here, we want to point out the non trivial dependence of the retarded self-energy on the 
parameters of the AC voltage. With this non trivial dependence the QD DOS strongly deviates from the 
usual single-particle Tien-Gordon behavior.\cite{Tien1,Tien2}


\begin{thebibliography}{12}
\bibitem{Fer}
As we discuss in the main text, our perturbation scheme in the on-site interaction 
overestimates $T_K$. The actual $T_K$ for the parameters we consider can be estimated by using 
$T_K\simeq\sqrt {2U\Gamma} e^{-\pi[|E_F -\epsilon _0|(U+\epsilon_0)
]/2\Gamma U}$, which gives
$T_{K}\approx 0.12\Gamma>T$.
Fermi liquid theory is, then, a good approximation
to describe the dynamical and transport properties of the QD.
\bibitem{Sat}
Note, however, that these satellites can only be observed by measuring the
differential conductance $dI/dV_{dc}|_{V_{dc}=\omega_0}$. The finite dc voltage is an extra source
of decoherence, i.e., reduction and broadening of the satellite peaks, \cite{Win,Gla2} not addressed here.
\end{thebibliography}
\end{document}